\documentclass{emulateapj}

\usepackage[usenames]{color}

\usepackage{rotating}

\shorttitle{How Many IRDCs form Massive Stars and Clusters?}
\shortauthors{Kauffmann \& Pillai}

\begin{document}

\title{How many Infrared Dark Clouds can form Massive
  Stars and Clusters?}

\author{Jens Kauffmann\altaffilmark{1,2} \&
  Thushara Pillai\altaffilmark{3,4}}

\altaffiltext{1}{Jet Propulsion Laboratory, California Institute of
  Technology, 4800 Oak Grove Drive, Pasadena, CA 91109, USA}
\altaffiltext{2}{NPP Fellow} \altaffiltext{3}{Caltech Astronomy
  Department, 1200 East California Blvd., Pasadena, CA 91125, USA}
\altaffiltext{4}{CARMA Fellow}

\slugcomment{Accelpted to The Astrophysical Journal Letters.}

\email{jens.kauffmann@jpl.nasa.gov}

\begin{abstract}
  We present a new assessment of the ability of Infrared Dark Clouds
  (IRDCs) to form massive stars and clusters. This is done by
  comparison with an empirical mass-size threshold for massive star
  formation (MSF). We establish
  $m(r)>870{}\,M_{\sun}\,(r/{\rm{}pc})^{1.33}$ as a novel approximate
  MSF limit, based on clouds with and without MSF. Many IRDCs, if not
  most, fall short of this threshold. Without significant evolution,
  such clouds are unlikely MSF candidates. This provides a first
  quantitative assessment of the small number of IRDCs evolving
  towards MSF. IRDCs below this limit might still form stars and
  clusters of up to intermediate mass, though (like, e.g., the
  Ophiuchus and Perseus Molecular Clouds). Nevertheless, a major
  fraction of the mass contained in IRDCs might reside in few $10^2$
  clouds sustaining MSF.
\end{abstract}


\maketitle

\section{Introduction}
About a decade ago, Galactic Plane surveys revealed large numbers of
Infrared Dark Clouds (IRDCs, \citealt{egan1998:irdcs};
\citealt{perault1996:isocam-galaxy}). These are identified as dark
patches against the diffuse Galactic mid-infrared background. First
studies of very opaque IRDCs suggested that these have very high
densities, column densities, and masses
($n[{\rm{}H_2}]\gtrsim{}10^5~\rm{}cm^{-3}$,
$N[{\rm{}H_2}]\gtrsim{}10^{23}~\rm{}cm^{-2}$,
$m\gtrsim{}10^3\,M_{\sun}$; \citealt{carey1998:irdc-properties}).
Since they are dark, they are likely to be in an early
  evolutionary phase. Embedded in IRDCs are ``cores'' of a few dozen solar masses
\citep{carey2000:irdc-submillimeter}. It has therefore been suggested
that many IRDCs are the long-sought examples of clouds just at
the onset of the formation of massive stars and (proto-)clusters. This
notion was corroborated by observations of young massive stars in a
few individual IRDCs (\citealt{rathborne2005:msf-irdc,
  rathborne2007:irdc-msf}; \citealt{pillai2006:g11};
\citealt{beuther2007:msf-irdc}). 
Such views also form the framework of schemes for IRDC evolution (e.g.,
\citealt{rathborne2006:irdcs-clusters},
\citealt{rygl2010:infrared-ext-clouds}) and reviews (e.g.,
\citealt{menten2005:irdc-review},
\citealt{beuther2007:massive-sf}). IRDC samples are usually compared
to regions of massive star formation (MSF), such as Orion and
M17 (e.g.,
\citealt{ragan2009:irdc-extinction}).\label{sec:introduction}

This picture cannot be complete, though. The above studies (and
\citealt{peretto2009:irdc-catalogue}) acknowledge that regions forming
low and intermediate mass stars can also appear as shadows in images
at mid-infrared wavelength \citep{abergel1996:ophiuchus}. Such IRDCs
will not form massive stars. Unfortunately, the number of IRDCs
evolving towards MSF is presently not known. Fractions up to 100\%
have been considered in the past (Section \ref{sec:most-stars-irdc}).

In this letter, we thus use a novel criterion to provide the first
conclusive quantitative demonstration that only few IRDCs are headed
towards MSF. This aids identifying pre-MSF IRDCs as targets for ALMA
and Herschel. As a bonus, the MSF threshold identified below---the
first observational limit of this kind---informs theory.

\begin{figure}
\includegraphics[width=\linewidth,bb=15 11 360 387,clip]{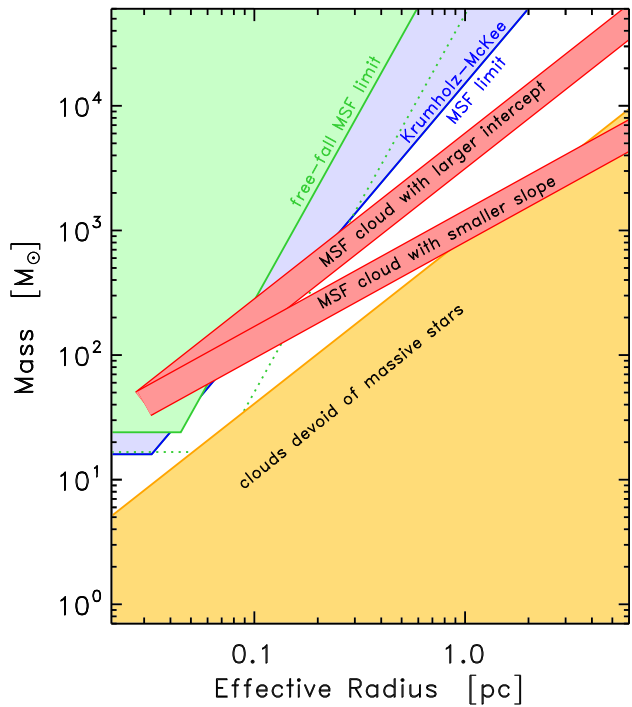}
\caption{Plausible theoretical mass-size limits for massive star
  formation (MSF, Section \ref{sec:msf-limits}; \emph{green and blue
    shading}) in relation to mass-size laws (e.g.,
    $m[r]=m_0\cdot{}r^b$) observed for non-MSF clouds (Eq.\
  \ref{eq:mass-size-limit}, Fig.\ \ref{fig:mass-size-comparison};
  \emph{yellow shading}). At small radii, MSF clouds (\emph{highlighted in
    red}) must contain fragments bound by the theoretical
  MSF-limits. Depending on the interplay of slope, $b$, and intercept,
  $m_0$, such clouds must also at radii $\gtrsim{}0.1~\rm{}pc$ be more
  massive than fragments in non-MSF clouds.\label{fig:msf-limit}}
\end{figure}

In papers I and II \citep{kauffmann2010:mass-size-i,
  kauffmann2010:mass-size-ii}, we show that solar neighborhood clouds
devoid of MSF (specifically: Perseus, Ophiuchus, Taurus, and Pipe
Nebula) generally obey
\begin{equation}
m(r)\le{}870\,M_{\sun}\,(r/{\rm{}pc})^{1.33} \, .
\label{eq:mass-size-limit}
\end{equation}
IRDCs submitting to Eq.\ (\ref{eq:mass-size-limit}) would resemble,
e.g., Ophiuchus and Perseus, but not Orion (which violates Eq.\
\ref{eq:mass-size-limit}). Figure \ref{fig:msf-limit} illustrates why
clouds bound for MSF must exceed Eq.\
(\ref{eq:mass-size-limit}). Since star formation necessitates an
appropriate mass reservoir, MSF requires that a large mass is
concentrated in a relatively small volume. Based on more detailed
theoretical considerations, Section \ref{sec:msf-limits} puts
quantitative limits on this intuitively evident reasoning. As seen in
Fig.\ \ref{fig:msf-limit}, the masses in this MSF region are well
above the mass--size range bound by Eq.\
(\ref{eq:mass-size-limit}). Observations of MSF clouds confirm Eq.\
(\ref{eq:mass-size-limit}) as a true MSF limit (Section
\ref{sec:msf-threshold}). This suggests to use Eq.\
(\ref{eq:mass-size-limit}) to roughly separate IRDCs with (future) MSF
from those without.\medskip

\noindent{}This letter is organized as follows. Based on data from Section
\ref{sec:method}, Section \ref{sec:msf-threshold} confirms (using
known MSF clouds) that Eq.\ (\ref{eq:mass-size-limit}) approximates an
MSF limit. Many well-studied IRDCs (25\%--50\%) fall short of this
threshold (Section \ref{sec:well-studied-irdcs}). Less certain data
for complete IRDC samples suggests that most IRDCs obey Eq.\
(\ref{eq:mass-size-limit}), and will thus not form massive stars
(Section \ref{sec:typical-irdcs}). Still, most of the mass contained
by IRDCs might be in clouds forming massive stars (i.e., those
violating Eq.\ \ref{eq:mass-size-limit}).

\section{Method \& Data}
\label{sec:method}

\subsection{Sample}
Data for solar neighbourhood clouds not forming massive stars (here:
Taurus, Perseus, Ophiuchus, Pipe Nebula) are taken from paper II (and
references therein). We rely on bolometer surveys to characterize MSF
sites: \citet{beuther2002:hmpo-densities} study FIR color-selected MSF
candidates with CS-detected dense gas but no radio continuum;
\citet{mueller2002:massive-sf} map water masers embedded in CS clumps
of high bolometric luminosity ($>10^3\,L_{\sun}$);
\citet{hill2005:simba-cold-cores} explore methanol masers and
ultra-compact H{\sc{}ii} regions; \citet{motte2007:cyg-x} study the
nearby Cygnus-X MSF site (we use their `clumps'). To exclude fragments
not forming massive stars, we only use the `Type 1' sources
($\rm{}CH_3OH$ and/or $\rm{}CH_3CN$ emission, no resolved radio
continuum) from the \citet{beuther2002:hmpo-densities} survey, and
ignore the secondary `mm-only' cores (without masers and H{\sc{}ii}
regions) in the \citet{hill2005:simba-cold-cores} study.

The IRDC samples were created using MSX and Spitzer
images. \citeauthor{rathborne2006:irdcs-clusters}
(\citeyear{rathborne2006:irdcs-clusters}; using bolometers) and
\citeauthor{ragan2009:irdc-extinction}
(\citeyear{ragan2009:irdc-extinction}; using $8~\mu{}\rm{}m$
extinction) focus on clouds with stark $8~\mu{}\rm{}m$
contrast. \citet{simon2006:irdc-characterization} report
$\rm{}^{13}CO$-based results for all IRDCs evident in their
$\rm{}^{13}CO$ Galactic Plane
survey. \citet{peretto2009:irdc-catalogue} catalogue extinction
properties for $11,000$ Spitzer $8~\mu{}\rm{}m$ IRDCs with unknown
distances.

\subsection{Data Processing}
\begin{figure}
\begin{tabular}{lc}
\includegraphics[width=0.75\linewidth,bb=15 53 360 387,clip]{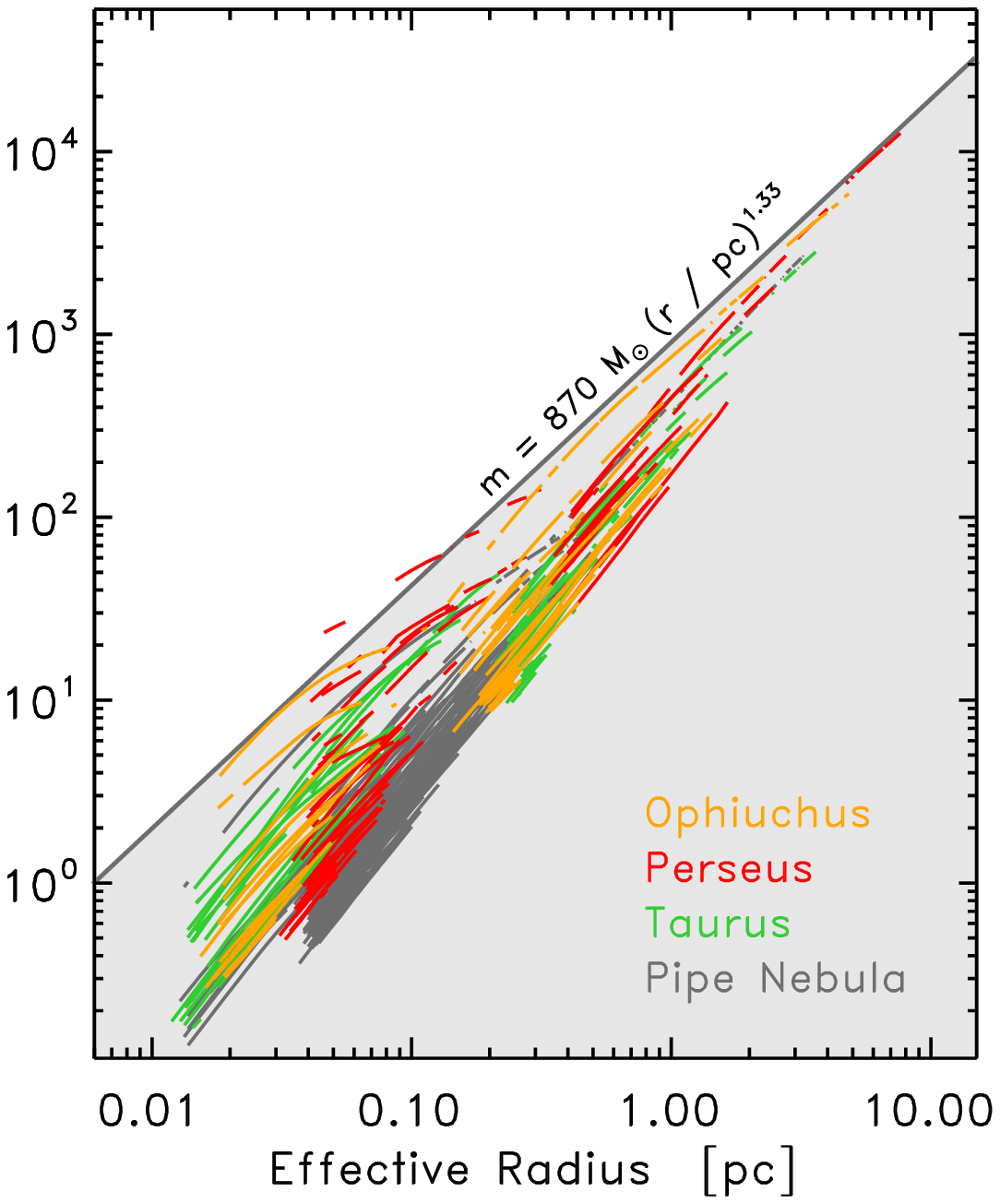} &
\begin{sideways}
\textsf{\hspace{1.8cm}\large{}(a) non-MSF clouds}
\end{sideways}\\
\includegraphics[width=0.75\linewidth,bb=15 53 360 387,clip]{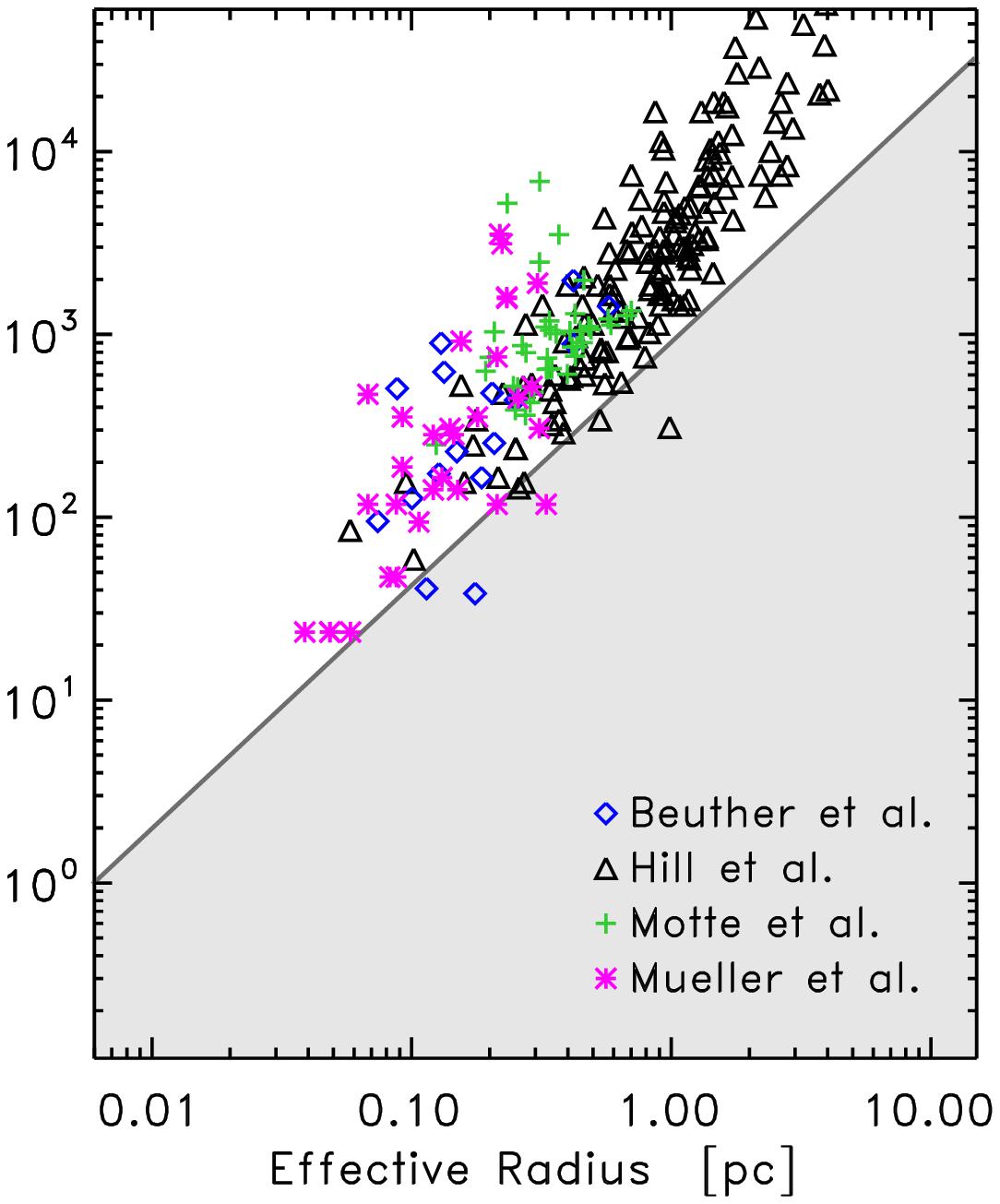} &
\begin{sideways}
\textsf{\hspace{1.8cm}\large{}(b) MSF clouds}
\end{sideways}\\
\includegraphics[width=0.75\linewidth,bb=15 11 360 387,clip]{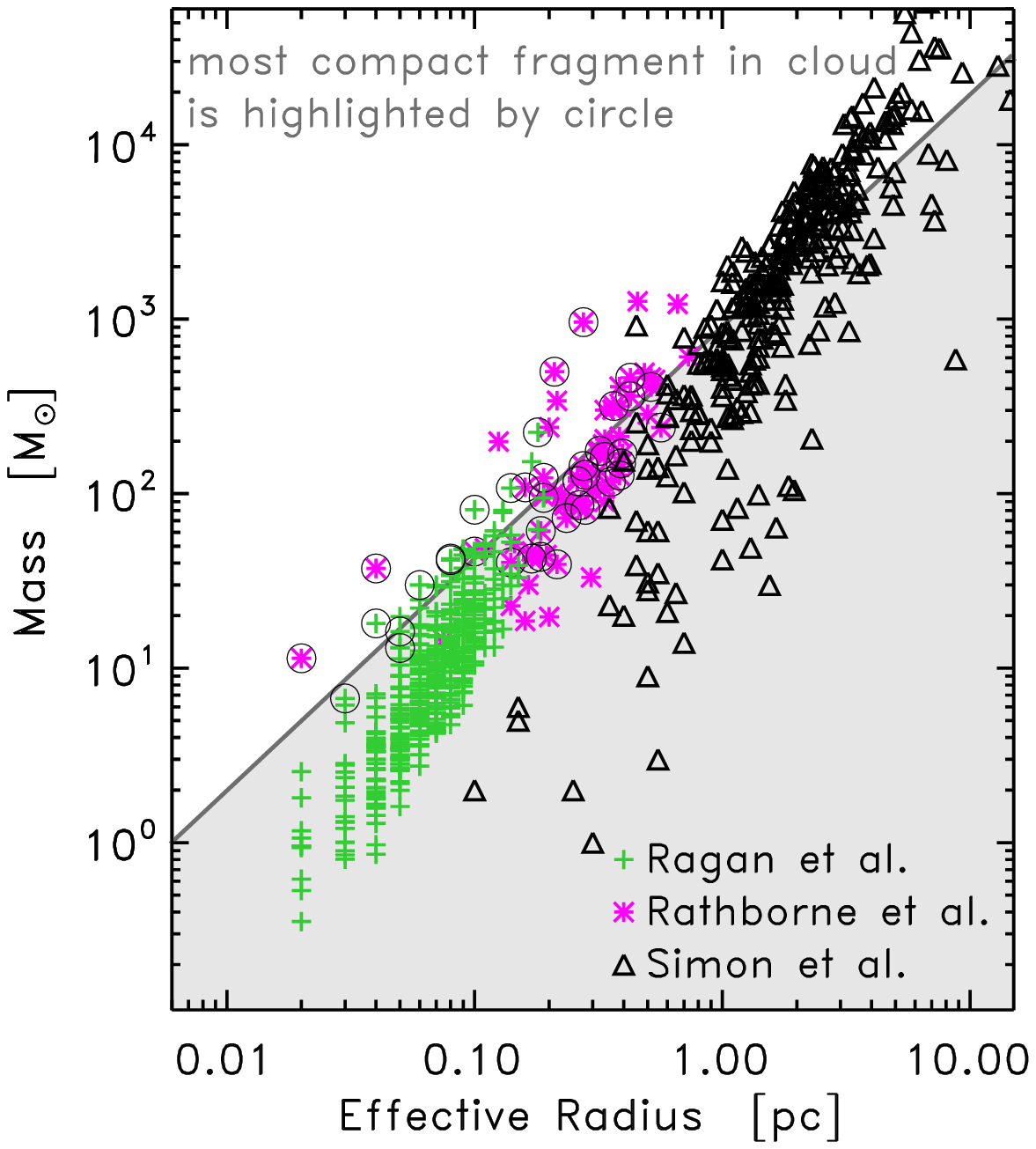} &
\begin{sideways}
\textsf{\hspace{2.5cm}\large{}(c) IRDCs}
\end{sideways}
\end{tabular}
\caption{Clouds with (\emph{panel b}) and without massive star
  formation (MSF; \emph{panel a}), compared to IRDCs (\emph{panel
    c}). Conceptually, data in panels (b) and (c) correspond to
  picking a single mass-size measurement along one of the lines shown
  in panel (a).  The MSF and non-MSF clouds suggest that the indicated
  limiting power law (Eq.\ \ref{eq:mass-size-limit}) approximates a
  mass-size limit for MSF (Section \ref{sec:msf-threshold}). Only a
  fraction of the IRDCs exceed this MSF limit (Fig.\
  \ref{fig:compactness}, Section \ref{sec:well-studied-irdcs}). If a
  star-forming region contains more than one fragment (i.e.,
    clump, core, etc.), the most compact fragment (i.e., with maximum
  $m[r]/m_{\rm{}lim}[r]$) is highlighted by a
  circle.\label{fig:mass-size-comparison}}
\end{figure}

The mass-size data for solar neighborhood clouds are derived in paper
II (using methods summarized in Section 2.1 and Fig.\ 1 of paper
I). They are based on column density maps derived from dust emission
(MAMBO and Bolocam) and extinction (2MASS)
data. Using a dendogram method
introduced by \citet{rosolowsky2008:dendrograms}, starting from a set
of local column density maxima, a given column density map is
contoured with infinitesimal level spacing. Every contour defines the
boundary of a cloud fragment. We derive the contour-enclosed mass and
the effective radius, $r=(A/\pi)^{1/2}$. Subsequent
contours/fragments are usually nested. This defines
relationships between cloud fragments, essentially yielding series of
mass-size measurements. In Fig.\ \ref{fig:mass-size-comparison}, such
series are drawn using continuous lines.

To derive column densities from the extinction maps, we assume that
column density and visual extinction are related by
$N_{\rm{}H_2}=9.4\times{}10^{20}~{\rm{}cm^{-2}}\,(A_V/{\rm{}mag})$
\citep{bohlin1978:av_conversion}. To combine dust emission and
extinction observations, they must be calibrated with respect to one
another. In practice, we use \citet{ossenkopf1994:opacities} dust
opacities (decreased by a factor 1.5, to match observed opacity laws;
Section 4.2 of paper I) for emission-based masses.\medskip

For comparisons, we must scale all masses to the column density laws
from paper II. Also, it is necessary to harmonize the different
definitions of mass and size. The scaled data are shown in Fig.\
\ref{fig:mass-size-comparison}.

Where relevant, we use dust temperatures suggested by the original
studies. However, we substitute our choice of dust opacities and the
aforementioned 1.5 scaling factor.  $\rm{}^{13}CO$ masses are directly
taken from \citet{simon2006:irdc-characterization}, since their
$\rm{}^{13}CO$-to-mass conversion law is in rough agreement with
(i.e., by factors of 1.1--2.0 larger than) the extinction-calibrated
ones derived by \citet{pineda2008:x-factor}. We assume that dust
emission at $1.2~\rm{}mm$ wavelength and optical depth at
$8~\rm{}\mu{}m$ wavelength are related by
$F_{\nu}^{\rm{}beam}=50~{\rm{}mJy}\,(11\arcsec{}~{\rm{}beam})^{-1}
\cdot{}\tau_{8~\rm{}\mu{}m}$ (Eq.\ 4 of
\citealt{peretto2009:irdc-catalogue}), and derive column densities
from these intensities (assuming dust at $15~\rm{}K$, and using the
1.5 scaling factor). We thus increase the
\citet{ragan2009:irdc-extinction} masses (from their case `A') by a
factor 1.47 (to account for their choice of opacities and molecular
weights)\footnote{We further correct their masses by factors
  $4/(\pi{}[(r/{\rm{}pc})\cdot{}206,265\arcsec{}/(d/{\rm{}pc})]^2/
  1\farcs2^2)$, where $d$ is distance, since pixels per beam (as
  erroneously adopted) have to be replaced by pixels per clump in Eq.\
  (5) of \citet{ragan2009:irdc-extinction}.}.

In many cases (\citealt{beuther2002:hmpo-densities},
\citealt{hill2005:simba-cold-cores},
\citealt{rathborne2006:irdcs-clusters}, \citealt{motte2007:cyg-x}),
the size listed in the original publication refers to the contour at
half peak intensity, while the mass measurement includes emission at
much lower levels. In these cases, we assume that the sources have a
near-Gaussian shape (just as explicitly assumed in many of the
original papers). For such sources, the mass contained in the half
peak column density contour is just a fraction $\ln(2)\approx{}0.69$
of the total mass (Eq.\ A.23 of \citealt{kauffmann2008:mambo-spitzer};
the area at half peak intensity is
$\pi{}[\theta_{\rm{}FHWM}/2]^2$). Thus we reduce the mass to a
fraction $\ln(2)$, and use half of the published FWHM size as the
effective radius. \citet{mueller2002:massive-sf} list masses for a
sphere, not an aperture, and so the mass (taken for the smaller of
their radii) has to be scaled up by a factor of order
$\pi{}/2{}\approx{}1.57$ (\citealt{kauffmann2008:mambo-spitzer}, Eq.\
13). If more than one distance is listed for a given object, we adopt
the smaller one (yielding a lower limit to $m[r]/m_{\rm{}lim}[r]$
derived below).

\section{Analysis}
\subsection{A Threshold for Massive Star
  Formation?\label{sec:msf-threshold}}
At given radius, a cloud fragment (i.e., clump, core, etc.) can
be compared against Eq.\ (\ref{eq:mass-size-limit}) by deriving the
mass ratio $m(r)/m_{\rm{}lim}(r)$ (where
$m_{\rm{}lim}[r]=870\,M_{\sun}\,[r/{\rm{}pc}]^{1.33}$), to which we
refer as the `compactness'. ``Secondary cores'' (only listed by
  \citeauthor{ragan2009:irdc-extinction} and
  \citeauthor{rathborne2006:irdcs-clusters}) are suppressed by
  characterizing star-forming regions (i.e., a given massive star, or
  an entire IRDC) by their most compact fragment,
  $\max[m(r)/m_{\rm{}lim}(r)]$.

\begin{figure}
\includegraphics[width=\linewidth]{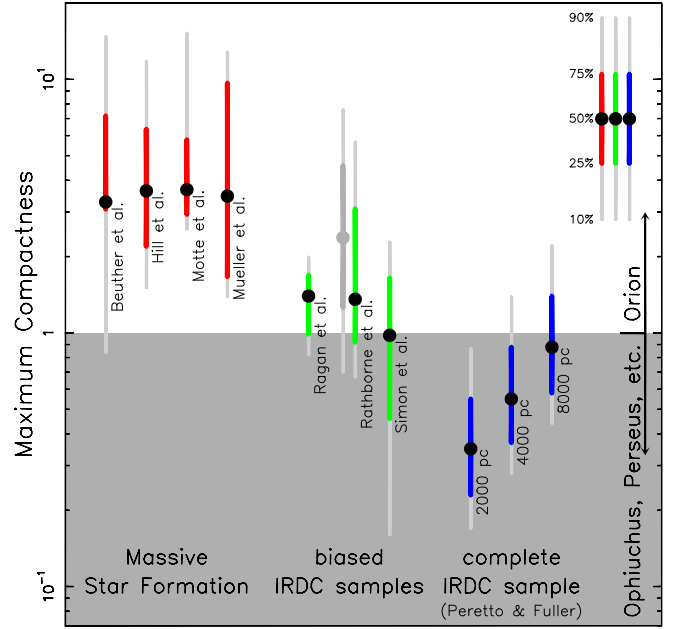}
\caption{Percentiles of the maximum compactness per cloud,
  $\max[m(r)/m_{\rm{}lim}(r)]$, for various cloud samples. For a given
  sample, the ratios below which a certain fraction (e.g., 25\%) of
  the sample members resides are indicated by bars. Local non-MSF
  clouds (Fig.\ \ref{fig:mass-size-comparison}[a]) have a compactness
  $\le{}1$ (Eq.\ \ref{eq:mass-size-limit}). The bars for the
  \citet{rathborne2006:irdcs-clusters} sample include (\emph{left}),
  respectively exclude (\emph{right}), their `em' cores. Clearly, the
  IRDCs do not reside in the mass-size space unambiguously associated
  with MSF.\label{fig:compactness}}
\end{figure}

Figure \ref{fig:compactness} gives $\max[m(r)/m_{\rm{}lim}(r)]$ as
derived for the samples examined here. This is based on the mass-size
data presented in Fig.\ \ref{fig:mass-size-comparison}. The
compactness assumes a range of values in every sample. This spread is
captured by plotting several percentiles.

As suggested by Fig.\ \ref{fig:mass-size-comparison}, we can clearly
see in Fig.\ \ref{fig:compactness} that regions forming massive stars
are, at given radius, more massive than the limiting mass,
$m_{\rm{}lim}(r)$. In all surveys of MSF regions,
$>75\%$ of the clouds have a maximum compactness $>1.7$. One survey
\citep{beuther2002:hmpo-densities} contains a very
small number of regions ($\sim10\%$) less compact than required by
Eq.\ (\ref{eq:mass-size-limit}). These regions might be interesting
targets for follow-up studies. In general, though, this analysis
corroborates the hypothesis that Eq.\ (\ref{eq:mass-size-limit})
approximates a threshold for MSF.

\subsection{Are IRDCs unusually Dense and
  Massive?\label{sec:well-studied-irdcs}}
Figure \ref{fig:compactness} provides a compactness analysis for
IRDCs. We separately characterize the
\citet{rathborne2006:irdcs-clusters} sample including and excluding
their `em' cores with associated $8~\mu{}\rm{}m$ sources (which are
not dark). ``True'' IRDCs will have properties in between these
extremes. Two interesting trends manifest in these
$m(r)/m_{\rm{}lim}(r)$ data.

First, IRDCs have masses which are, for given size, comparable to
those of solar neighborhood clouds not forming massive stars (e.g.,
Ophiuchus and Perseus). In all samples, $\gtrsim25\%$
of all clouds have a compactness $<1$. Except for the
\citet{rathborne2006:irdcs-clusters} clouds, $\ge75\%$ of all targets
exceed Eq.\ (\ref{eq:mass-size-limit}) by a factor $<2$.

Second, IRDCs are less compact than regions forming massive stars. For
example, excluding the \citet{rathborne2006:irdcs-clusters} targets,
$>75\%$ of all IRDCs are less compact than most ($>75\%$) of the
MSF regions.

In summary, the IRDCs studied here have (for given size) masses in
between those of regions with and without MSF (where ``true''
  \citeauthor{rathborne2006:irdcs-clusters} IRDCs have properties in
  between the two extremes shown). Very clearly, they do not reside
in the mass-size space unambiguously associated with the formation of
massive stars.\medskip

However, before drawing final conclusions, let us consider some biases
affecting our analysis. First, \citet{ragan2009:irdc-extinction}
derive masses using CLUMPFIND, while
\citet{rathborne2006:irdcs-clusters} use GAUSSCLUMPS. For the former,
paper I showed explicitly that the derived masses are, for given
radius, just $\lesssim70\%$ of those derived
using our dendrogram approach. For the latter, the same is expected, since the
Gaussian fits only describe a fraction of the emission. In a given
map, our characterization scheme from papers I and II would thus find
larger masses.

These biases are countered by other factors, though. We use the `case
A' masses (assuming bright IR foregrounds) provided by
\citet{ragan2009:irdc-extinction}. Following
\citet{peretto2009:irdc-catalogue}, their `case B' (fainter
  foregrounds) appears to be more realistic. The masses could thus be
lower by a factor $\sim2$
\citep{ragan2009:irdc-extinction}. Similarly,
\citet{pineda2008:x-factor} suggests $\rm{}^{13}CO$-to-mass conversion
factors lower than used by \citet{simon2006:irdc-characterization}. In
any case, similar biases affect the data for MSF regions. Differences
between these and IRDCs are not likely to only come from observational
uncertainties.

Finally, none of the IRDCs in the
\citeauthor{ragan2009:irdc-extinction},
\citeauthor{rathborne2006:irdcs-clusters}, and
\citeauthor{simon2006:irdc-characterization} samples are ``typical''
for the general Galactic population. \citet{ragan2009:irdc-extinction}
and \citet{rathborne2006:irdcs-clusters} select clouds which are
unusually dark in $8~\rm{}\mu{}m$
images. \citet{simon2006:irdc-characterization} only characterize
IRDCs which are relatively large and dark, and are clearly detected in
$\rm{}^{13}CO$ emission. All this excludes IRDCs of low mass and
density from the samples. Less biased IRDC samples should thus be less
compact than derived here.

\subsection{Typical Star Formation Properties in the
  Galaxy\label{sec:typical-irdcs}}
The \citet{peretto2009:irdc-catalogue} catalogue lists IRDC angular
sizes and column densities for the entire Galactic Plane covered by
Spitzer. It thus provides an ideal tool to derive a first idea of
typical IRDC properties. Since they likely constitute (to our present
knowledge) the typical reservoir of Galactic star-forming gas, IRDC
characteristics probably gauge the early state of
Galactic star forming regions.

\begin{figure}
\includegraphics[width=\linewidth,bb=15 11 360 387,clip]{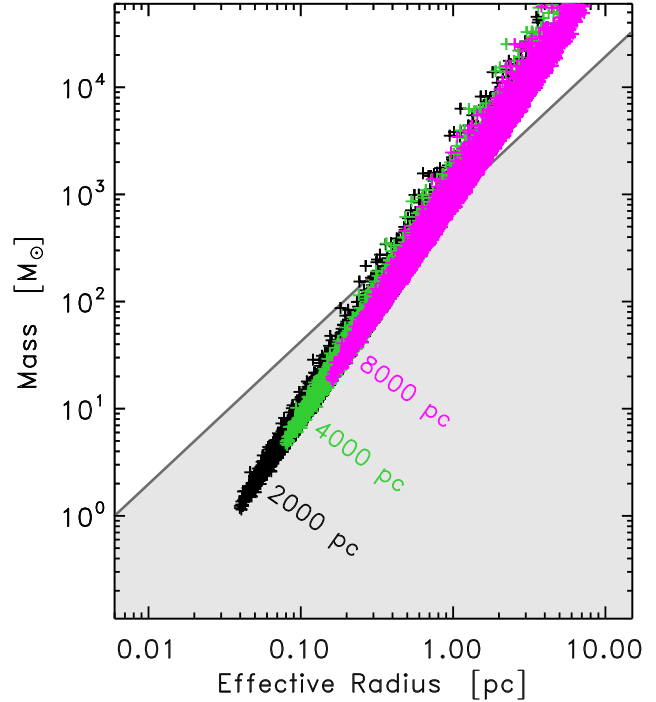}
\caption{Like Fig.\ \ref{fig:mass-size-comparison}(c), but for
  the \citet{peretto2009:irdc-catalogue} sample (projected out to
  various distances).\label{fig:peretto-fuller}}
\end{figure}

\begin{table}
\caption{Compact \citet{peretto2009:irdc-catalogue}
  IRDCs\label{tab:peretto-fuller}}
\begin{tabular}{cccccccccccccc}
\hline\hline
\rule[-1.5ex]{0ex}{5.0ex}Distance$^a$&Number$^b$&Fraction$^c$&
  Mass$^b$&Mass Fraction$^c$\\
\rule[-1.5ex]{0ex}{0ex}kpc&--&\%&$10^6\,M_{\sun}$&\%\\
\hline
\rule[0ex]{0ex}{3ex}2&831&7&2.0&71\\
4&2218&20&9.8&87\\
6&3639&32&23.6&93\\
\rule[-1.5ex]{0ex}{0ex}8&4778&42&43.2&96\\
\hline
\end{tabular}

\rule[0ex]{0ex}{3ex}$^a$distance to which the sample is projected

$^b$number of clouds with $m(r)>m_{\rm{}lim}(r)$, and their total mass

$^c$mass and number fraction of compact clouds
\end{table}

Since no distances are known for the
  \citeauthor{peretto2009:irdc-catalogue} IRDCs,
we constrain their masses and sizes assuming a
reasonable range of distances. Analysis by
\citet{simon2006:irdc-characterization} and
\citet{jackson2008:irdc-distances} suggests that most IRDCs have
distances of $2$--$8~\rm{} kpc$. Figure \ref{fig:peretto-fuller}
illustrates the derived masses and sizes, and Table
\ref{tab:peretto-fuller} characterizes the IRDCs found to be compact
(i.e., $m[r]/m_{\rm{}lim}[r]>1$).

This analysis has two interesting results. First, by number, most of
the \citet{peretto2009:irdc-catalogue} IRDCs have masses and sizes
comparable to those of solar neighborhood clouds devoid of massive
stars (i.e., they are not compact). This holds even when adopting the
largest reasonable distance. Second, the compact clouds contain most
of the mass (more accurately: most of the area-integrated column
density) seen in these IRDCs, even for small IRDC distances.

Unfortunately, the \citet{peretto2009:irdc-catalogue} survey is (like
most extinction studies) uncertain in the sense that it assumes that
the diffuse Galactic emission can be reliably modelled in its spatial
distribution. This may not be true. In this spirit, the results from
this section should be taken as an indication, not as a final result.

\section{Discussion}
\subsection{Mass-Size Structure of MSF Clouds\label{sec:msf-limits}}
Consider the following toy model to understand the expected mass-size
properties of MSF clouds. Stars probably form on
a timescale $\tau_{\rm{}sf}$ slower ($\varepsilon_{\rm{}ff}<1$) than the
free-fall timescale\footnote{$\tau_{\rm{}ff} =
(3\pi{}/[32\,G\,\langle{}\varrho{}\rangle])^{1/2}$, where
$G$ is the constant of gravity and $\langle{}\varrho{}\rangle$ is the
volume-averaged density},
$\tau_{\rm{}sf}\gtrsim{}\tau_{\rm{}ff}/\varepsilon_{\rm{}ff}{}\propto
1/(\varepsilon_{\rm{} ff}\,\langle{}\varrho{}\rangle^{1/2})$.
In spherical symmetry, mass, size, and density are
related by $\langle{}\varrho{}\rangle=
\varepsilon_{\varrho}\,m/(4/3\,\pi{}r^3)$, where
$\varepsilon_{\varrho}<1$ takes line-of-sight material not
associated with the sphere into account. A specific star formation
timescale then requires that
\begin{equation}
m(r)\gtrsim{}27.1\,M_{\sun}\,
\frac{1}{\varepsilon_{\varrho}\,\varepsilon_{\rm{}ff}^2}
\left(\frac{\tau_{\rm{}sf}}{10^5~\rm{}yr}\right)^{-2}
\left(\frac{r}{0.1~\rm{}pc}\right)^3 \, .
\label{eq:msf-ff}
\end{equation}
Further, to form a star of certain mass, $M_{\star}$, a mass reservoir
larger than $M_{\star}$ is necessary:
\begin{equation}
m(r)\gtrsim{}M_{\star}/\varepsilon_m\,.
\label{eq:msf-efficiency}
\end{equation}
Figure \ref{fig:msf-limit} evaluates these limits for a star of
$8\,M_{\sun}$, based on efficiencies\footnote{The main accretion phase
  of a low mass star (IR-classes 0 and I) typically finishes after
  $7\times{}10^5~\rm{}yr$ \citep{evans2009:c2d-summary}. Typical
  free-fall timescales of their natal cores $\sim{}10^5~\rm{}yr$
  \citep{enoch2008:lifetimes_and_cmf} then imply
  $\varepsilon_{\rm{}ff}\approx{}1/7$. Further,
  $\varepsilon_{\varrho}\approx{}1/2$
  (\citealt{kauffmann2008:mambo-spitzer}, Eq.\ 13) and
  $\varepsilon_m\approx{}1/3$ \citep{alves2007:imf}. Since massive
  stars might form faster, and the star formation efficiency is not
  constrained well, we explore
  $3\le{}\tau_{\rm{}sf}/10^5~{\rm{}yr}\le{}7$ and
  $1/3\le{}\varepsilon_m{}\le{}1/2$ in Fig.\ \ref{fig:msf-limit}} and
timescales from Spitzer observations of solar neighborhood
clouds. Within the model, cloud collapse will only yield a massive
star if initiated inside the boundaries set by Eqs.\
(\ref{eq:msf-ff}--\ref{eq:msf-efficiency}).
\citet{krumholz2008:column-density-threshold} provide a similar limit,
derived assuming that the collapsing region is heated by a
  cluster of low-mass stars (in our
terminology, they use $\varepsilon_m=1/2$).

In order to sustain MSF, at least a few cloud fragments in MSF clouds
must reside within the theoretical MSF boundaries mentioned above
(Fig.\ \ref{fig:msf-limit}). The global structure of these clouds can
usually be described by power laws, $m(r)=m_0\cdot{}r^b$, with $b<2$
(paper II). Such power laws imply that MSF clouds violate
$m(r)<m_{\rm{}lim}(r)$ (Eq.\ \ref{eq:mass-size-limit}). Depending on
slope ($b$), intercept ($m_0$), and their interplay, such excesses are
expected for radii $\gg{}0.1~\rm{}pc$ (Fig.\
\ref{fig:msf-limit}). This is just what we find for MSF clouds (Fig.\
\ref{fig:mass-size-comparison}[b]).

MSF is thus only possible if a clouds' slopes are shallow, intercepts
are large, or both, when compared to Eq.\
(\ref{eq:mass-size-limit}). This permits a new way to quantitatively
compare the structure of clouds with and without MSF. Pure differences
in $m_0$ imply that MSF and non-MSF clouds only differ in their
absolute properties. Differences in slopes $b$, however, imply
\emph{relative} differences in the structure, such as deviations in
the hierarchical cloud structure.

\subsection{Average State of IRDCs}
The IRDC properties mentioned in the introduction
($n[{\rm{}H_2}]\gtrsim{}10^5~\rm{}cm^{-3}$,
$N[{\rm{}H_2}]\gtrsim{}10^{23}~\rm{}cm^{-2}$,
$m\gtrsim{}10^3\,M_{\sun}$) only seem to characterize the densest
patches in very large and massive IRDCs. They are not well suited to
describe IRDCs on average.

Some IRDCs with $m(r)<m_{\rm{}lim}(r)$ might further evolve and
eventually undergo MSF. And particular dust properties could, in
principle, erroneously indicate $m(r)<m_{\rm{}lim}(r)$ where the
reverse is true. However, such caveats are not usually considered when
using IRDC data to constrain MSF. Thus we abstain from such
considerations.

Our study suggests that many IRDCs, if not most, are not related to
MSF. One thus has to be prudent when using IRDC properties to
constrain MSF initial conditions. Most studies discussing IRDCs as
pre-MSF sites concentrated on very opaque IRDCs of large angular
size. These clouds often violate Eq.\ (\ref{eq:mass-size-limit}), and
many of them are good MSF candidates.

\subsection{Do most stars form in just few
  IRDCs?\label{sec:most-stars-irdc}}
\citet{rathborne2006:irdcs-clusters} suggest that most of the Galactic
star formation might come from IRDCs. The absence of other likely
reservoirs of star-forming gas evinces this too. By number, most IRDCs
are likely to form stars and clusters of low and intermediate mass,
just as Ophiuchus and Perseus do.

Still, many IRDCs will turn towards MSF. Interestingly, Table
\ref{tab:peretto-fuller} suggests that most of the mass located in
IRDCs is in clouds that will form massive stars. For example, the 250
most compact clouds from the \citet{peretto2009:irdc-catalogue} sample
(identified assuming a common distance) contain more than 50\% of the
area-integrated column density of all IRDCs. This suggests that they
also contain a major fraction of the mass seen in IRDCs. If this
reasoning is correct, just few $10^2$ IRDCs (and not all $\sim{}10^4$:
\citealt{rathborne2006:irdcs-clusters}) might contain most of the
Galaxy's star-forming gas. Given the uncertain nature of the
properties derived from the \citet{peretto2009:irdc-catalogue} data
(Section \ref{sec:typical-irdcs}), this conclusion is far from
certain, though.

\section{Conclusions}
This letter studies whether Infrared Dark Clouds (IRDCs) are able to
form massive stars. Our main conclusions are as follows.
\begin{itemize}
\item Observations of regions with and without massive star formation
  (MSF) suggest that the condition
  $m(r)\le{}870\,M_{\sun}\,(r/{\rm{}pc})^{1.33}$ (Eq.\
  \ref{eq:mass-size-limit}) approximates a threshold for MSF (Section
  \ref{sec:msf-threshold}). MSF clouds differ from those obeying Eq.\
  (\ref{eq:mass-size-limit}) in mass-size slope or intercept (Fig.\
  \ref{fig:msf-limit}, Section \ref{sec:msf-limits}).
\item Many IRDCs (Section \ref{sec:well-studied-irdcs}), if not most
  (Section \ref{sec:typical-irdcs}), do not exceed Eq.\
  (\ref{eq:mass-size-limit}). Without significant further evolution,
  such clouds are unlikely candidates for MSF, but they might well
  form stars and clusters of up to intermediate mass (like Perseus and
  Ophiuchus). Very opaque IRDCs of large angular size
  constitute good MSF candidates.
\item Provided extinction-based masses can be trusted, just few $10^2$
  IRDCs might contain a major fraction of the Galaxy's star-forming
  gas (Section \ref{sec:most-stars-irdc}). These IRDCs would be dense
  and massive enough to host MSF.
\end{itemize}

\acknowledgements{We are indebted to a careful referee, who
  significantly helped to improve the text. This research was
  supported by an appointment of Jens Kauffmann to the NASA
  Postdoctoral Program at the Jet Propulsion Laboratory, administered
  by Oak Ridge Associated Universities through a contract with NASA.}


\end{document}